\documentclass[showpacs,preprintnumbers,amsmath,amssymb,showkeys,nofootinbib,prb,twocolumn]{revtex4}

\usepackage{graphicx}
\usepackage{dcolumn}
\usepackage{bm}
\usepackage{textcomp}
\usepackage[letterpaper,left=2.0cm,right=1.2cm,bottom=4.5cm,top=2cm]{geometry}

\usepackage{hyperref}

\newcommand{\thermal}{\phi}
\newcommand{\pure}{\pi}
\newcommand{\pureBar}{\bar\pi}
\hypersetup{colorlinks=true,citecolor=blue}
\newlength{\mywidth}
\begin{document}

\title{Production of minimally entangled typical thermal states\\
 with the Krylov-space approach}

\author{G. Alvarez}
\affiliation{Computer Science \& Mathematics
Division and Center for Nanophase Materials Sciences, Oak Ridge National Laboratory,
 \mbox{Oak Ridge, TN 37831}, USA}

\date{\today}

\begin{abstract}
The minimally entangled typical thermal states algorithm is applied to fermionic systems
using the Krylov-space approach to evolve the system in imaginary time. The convergence of local observables
is studied in a tight-binding system with
a site-dependent potential. The temperature dependence
of the superconducting
correlations of the attractive Hubbard
model is analyzed on chains, showing an exponential
decay with distance and exponents proportional to the temperature at low temperatures, as expected.
In addition, the non-local parity correlator is calculated at finite temperature.
Other possible applications of the minimally entangled typical thermal states algorithm to fermionic
systems are also discussed.
\end{abstract}

\pacs{
02.70.-c,
03.67.Lx,
71.10.Fd,
74.25.Dw
}
\keywords{DMRG,time evolution,time-step targeting, METTS, temperature dependence}

\maketitle

\section{Introduction}
Originally, the density matrix renormalization group\cite{re:white92,re:white93} (DMRG) algorithm
dealt with zero temperature or ground-state properties.
The development of the algorithm at finite temperature has been a topic of much
interest,\cite{re:verstraete04,re:zwolak04,re:feiguin05}
because of the increased complexity associated with \emph{efficiently} computing temperature-dependent properties.

Recently, White\citep{re:white09} proposed an efficient method---that he refers to as
\emph{minimally entangled typical thermal states}
or METTS---to compute temperature-dependent observables with a complexity similar to
ground state computations.
An observable $\hat{A}$ of a quantum many body system at temperature $T=1/(k_B\beta)$ is expressed as
$\langle A\rangle={\rm Tr}{[}\hat{\rho}\hat{A}{]}=1/\mathcal{Z}{\rm Tr}{[}{\rm e}^{-\beta \hat{H}} \hat{A}{]}$,
the trace involving two kinds of integration: over quantum and thermal fluctuations.
Performing this calculation directly is intractable, even more so than at zero temperature.
One can, however, approximate the expectation value of $A$ by strategies based on sampling.
The METTS algorithm\cite{re:white09} starts from classical product states, and
then entanglement is brought about by the (imaginary) time evolution of those initial states.

In METTS, thermal fluctuations are sampled by randomly selecting quantum states.
To understand how this is done let us first expand the trace in terms of an orthonormal basis $|i\rangle$, such that
$\langle A\rangle=
\frac{1}{\mathcal{Z}}\sum_i P(i)\langle\phi(i)|A|\phi(i)\rangle,$
where $|\phi(i)\rangle=P(i)^{-1/2}{\rm e}^{-\beta H /2 }|i\rangle$ and $P(i)=\langle i| e^{-\beta H /2} |i\rangle$.
A choice for the basis $|i\rangle$ is the set of classical product states (CPS); these are states with wavefunctions
$|i\rangle=|i_0\rangle|i_1\rangle\cdots|i_{N-1}\rangle$, where the labels 0, 1, 2, \ldots, refer to the sites of the lattice.
The essence of the method lies in the way $\langle A \rangle$ is estimated by sampling the states
$|\phi(i)\rangle$
with probability $P(i)/\mathcal{Z}$, and by averaging the expectation values $\langle\phi(i)|A|\phi(i)\rangle$ computed at each step.
A proof that the ensemble $\{|\phi(i)\rangle\}$ so generated correctly reproduces all thermodynamic measurements
is given in reference~\onlinecite{re:white09}, as well as a justification for referring to the set $\{|\phi(i)\rangle\}$ as METTS.
 In addition to the original reference,\cite{re:white09} a matrix-product-state formulation of the method can
 be found in references~\onlinecite{re:stoudenmire10,re:schollwock10}, as well as a
viewpoint in Physics.\citep{re:schollwock09}

Quantum Monte Carlo methods can also simulate strongly correlated electron models at finite temperature.
The use of finite-temperature DMRG has, however, advantages and disadvantages that make it an ideal
complementary method.  Finite-temperature DMRG is advantageous in the case of long chains, and even ladder geometries.
Moreover, for those strongly correlated electron models where the sign problem hinders quantum Monte Carlo simulations,
DMRG methods can come to the rescue. They might be the only unbiased technique applicable to models with,
for example, spin flipping terms, which are known to present serious sign problems.
This is exactly the case of iron-\cite{re:kamihara08} and selenide-based\cite{re:guo10} superconductors, which
include J terms\cite{re:daghofer08,re:moreo08}
 (as in the case of t-J models). Thus, the METTS algorithm might be particularly appropriate for these superconductors.

This paper explains in detail the production and use of METTS with the Krylov-space approach for DMRG time
evolution\cite{re:manmana05,re:batrouni06,re:schollwock05}.
Section \ref{sec:algorithm} describes the implementation of the algorithm, including the ``collapse'' procedure,
the ergodicity issues, and the computational complexity.
Section \ref{sec:results} focuses on local observables in the case of a fermionic non-interacting system with a site-dependent potential.
The attractive Hubbard model at quarter filling is then studied for a one-dimensional open geometry (chain), and
superconducting correlations are compared to known results. The non-local parity correlator is then calculated at finite temperature.
Finally, section \ref{sec:summary} presents an outlook for the further applicability of METTS.

\section{Algorithm}\label{sec:algorithm}

\subsection{Production of METTS}

Here the Krylov-space approach for time evolution is adapted to produce minimally entangled typical thermal states or METTS,\cite{re:white09,re:stoudenmire10} that is,
a ``thermal'' evolution is produced. The natural real-space basis of a model is defined as the set of states
in its one-site Hilbert space. For example, the natural basis is composed of the states empty, up, down, and doubly occupied in the case of the
one-orbital Hubbard model. The notation $\mathcal{N}{|x\rangle\equiv|x\rangle/|||x\rangle||}$ appears below to simplify the expressions.

The steps to obtain observables at any target inverse temperature $\beta_{\rm T}$ are as follows.

(1) Set the current inverse temperature $\beta_{\rm C} = 0$. Do a standard ``infinite DMRG,'' and grow the $N-$site lattice,
choosing a random real-space basis state
per site to create a pure state.
Target this pure state at each step, that is, include it in the reduced density matrix.
 Proceed in this way until all sites have been added and
end up with a pure state over the whole lattice:
$|\pure\rangle = |i_0\rangle |i_1\rangle ... |i_{N-1}\rangle$.

(2) Obtain the states $|\thermal(k)\rangle = \mathcal{N}{\exp(-\beta_k H /2 ) |\pure\rangle}$, $0\le k<l$, using
a Krylov-space approach for the time
evolution;\cite{re:manmana05,re:batrouni06,re:schollwock05}
an implementation can be found in reference~\onlinecite{re:alvarez11}. Here
$\beta_k = k \tau/(l-1)$, $\tau=0.1$, $l=5$. Collapse (the collapse is described in section~\ref{sec:collapse}) the last one, $|\thermal(l-1)\rangle$,
into a pure state $|\pure'\rangle$.
Target $|\thermal(k)\rangle$  $0\le k < l$, and  $|\pure'\rangle$.

(3) Move the center of orthogonality by one, as ``finite'' DMRG does. Wave-function transform  $|\thermal(k=0)\rangle$ (also denoted by $|\pure\rangle$)
into ${\rm Wft}(|\thermal(k=0)\rangle)$. Recompute $\exp(-\beta_k H /2 ) {\rm Wft}( |\thermal(k)\rangle)$
 $\forall\,1\le k<l$.
Wave-function transform also the collapsed state $|\pure'\rangle$.
Proceed sweeping the lattice for a while, until, for example, an extreme is reached, or all sites have been visited at least once.

(4)
If $\beta_{\rm C}<\beta_{\rm T}$ then advance in $\beta$: Set $|\thermal(k=0)\rangle\equiv|\pure\rangle$ to $|\thermal(k=l-1)\rangle$.
Increase $\beta_{\rm C}$ by $\tau$. Go to step (2).
If $\beta_{\rm C} = \beta_{\rm T}$ then perform a measurement (for production runs, instead of
measuring \emph{in situ}, save the METTS to measure post-processing)  using the current wave-function transformed
$|\thermal(k=0)\rangle$.
Set the state $|\thermal(k=0)\rangle$ to the wave-function transformed collapsed $|\pure'\rangle$. In other words, set $|\pure\rangle$ to $|\pure'\rangle$.
Set $\beta_{\rm C}=0$ and go to step (2).

\subsection{Collapsing METTS}\label{sec:collapse}

Let us consider first the collapse into the natural basis and then into a random basis.

If the collapse happens in the real-space basis then for a Hubbard model with only one orbital there are four states to collapse
into: empty, up, down, and doubly occupied.
If  $|\thermal(k)\rangle$ (or its wave-function transformed form) is centered on site $i$ then
 $|\thermal(k)\rangle=\sum_{\alpha_L,\alpha_i,\alpha_R} A_{\alpha_L,\alpha_i,\alpha_R}|\alpha_L\rangle|\alpha_i\rangle|\alpha_R\rangle$, where
$\alpha_i$ is a state of the natural real-space basis.
This state is normalized, hence $\sum_{\alpha_L,\alpha_i,\alpha_R} |A_{\alpha_L,\alpha_i,\alpha_R}|^2=1$.

Let $|\pure(\alpha_i)\rangle=\sum_{\alpha_L,\alpha_R} A_{\alpha_L,\alpha_i,\alpha_R}|\alpha_L\rangle|\alpha_i\rangle|\alpha_R\rangle$,
for each state $\alpha_i$ of the
natural one-site basis at $i$. Let $p(\alpha_i)=||\pure(\alpha_i)\rangle||^2$. The condition $\sum_{\alpha_i} p(\alpha_i)=1$ follows
from the normalization of $|\thermal(k)\rangle$.
A state $\alpha_i$  is selected with probability $p(\alpha_i)$ and the collapse occurs into the state
$|\pure'\rangle=\mathcal{N}|\pure(\alpha_i)\rangle$, which is now to be used for step (2).

When the collapse happens in a random basis defined by $|\alpha_i\rangle = \sum_{\eta_i} M_{\alpha_i,\eta_i}|\eta_i\rangle$ we proceed as follows.
First we rewrite
\begin{equation}
|\thermal(k)\rangle=\sum_{\alpha_L,\alpha_i,\alpha_R} A_{\alpha_L,\alpha_i,\alpha_R} \sum_{\eta_i} M_{\alpha_i,\eta_i}|\alpha_L\rangle|\eta_i\rangle|\alpha_R\rangle.
\end{equation}
Defining
 $|\pureBar(\eta_i)\rangle$ by $|\thermal(k)\rangle=\sum_{\eta_i} |\pureBar(\eta_i)\rangle$,
for each state $\eta_i$ of the
random basis, yields
\begin{equation}
|\pureBar(\eta_i)\rangle=\sum_{\alpha_L,\alpha_i,\alpha'_i,\alpha_R} M^{-1}_{\eta_i,\alpha'_i} A_{\alpha_L,\alpha_i,\alpha_R} M_{\alpha_i,\eta_i}
|\alpha_L\rangle|\alpha'_i\rangle|\alpha_R\rangle.
\end{equation}
The new state $\mathcal{N}\pureBar(\eta_i)$ is collapsed with probability  $p(\eta_i)=||\pureBar(\eta_i)\rangle||^2$;
these probabilities add up to 1 because of normalization.

One important practical consequence of these collapse equations is that they do
not preserve local symmetries unless the collapse basis does, implying that simulations, in most cases, will have to be performed in the \emph{grand canonical ensemble}.
Indeed, all METTS simulations in this paper are done in the grand canonical ensemble, as we will see in the next sections.

\subsection{Ergodicity}
To illustrate ergodicity issues with this technique let us
consider a tight-binding chain and a chemical potential term such that
$H=\sum_{i,j}t_{ij} c^\dagger_{i}c_{j} +  \mu\sum_{i} n_{i}$, where $t_{ij}=1$ if  $i$ and $j$ are
nearest neighbors and $0$ otherwise.
As observables let us consider the total energy and density of the system at inverse temperature $\beta$.
These observables depend on $\mu$, but  if we use the natural basis for the collapse, the
method will (incorrectly) yield values independent of $\mu$.
To understand the reason for the second statement I would now like to show that, \emph{if we
collapse only to the natural basis,} METTS do not depend on $\mu$.
Let us choose an initial CPS $|\pi\rangle$, and compute the corresponding METTS at $\beta$.
Note that $|\pi\rangle$ is an eigenvector of $\mu\sum_{i} n_{i}$.
The resulting METTS is $\exp(-\beta H_0) |\pi\rangle / \langle \pi| \exp(-\beta H_0) |\pi\rangle$, where
$H_0=\sum_{i,j}t_{ij} c^\dagger_{i}c_{j}$ does not contain $\mu$, which canceled out due to normalization.
 If we now proceed to collapse this METTS into
a CPS \emph{in the natural} basis, we obtain the CPS $|\pi'\rangle$. Evolving $|\pi'\rangle$ does not involve $\mu$,
as $|\pi'\rangle$ is an eigenvector of $\mu\sum_{i} n_{i}$.
Therefore, all the METTS thus obtained are independent of $\mu$ and, when measuring over them, we will (incorrectly) obtain
values independent of $\mu$.

\begin{figure}
\includegraphics[width=\mywidth,clip]{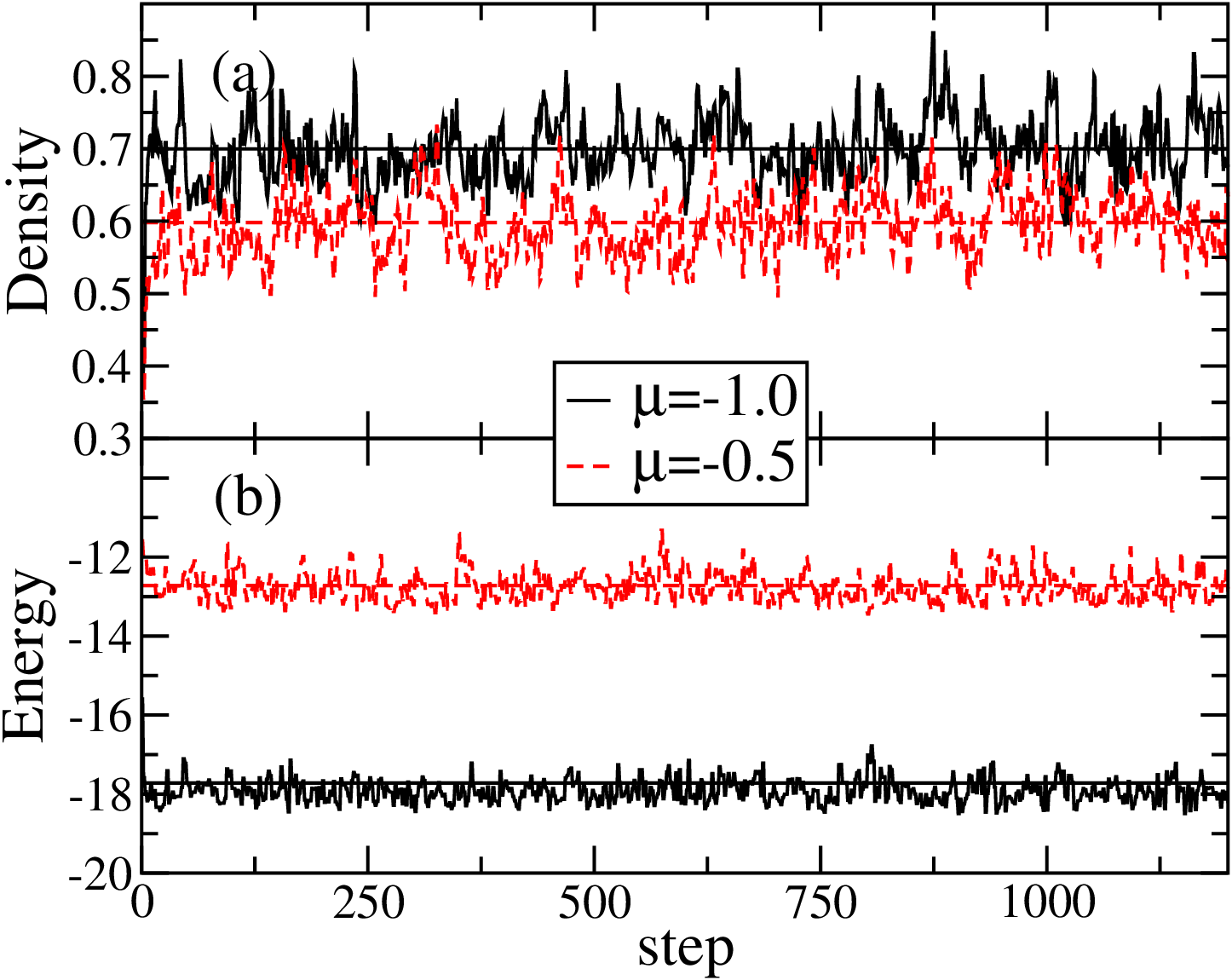}
\caption{(Color online) (a) Density and (b) energy of electrons in a tight-binding chain with a chemical potential versus the step
of the METTS algorithm. Here METTS are collapsed in a random basis, reducing ergodicity issues, and causing
the observables' average to tend to their exact values (horizontal lines).\label{fig:figure1mu}}
\end{figure}

In this case the solution is simple: the collapse must be carried out into a random basis and not into the natural basis.
Results are shown in figure \ref{fig:figure1mu}. In general, the Hamiltonian can be decomposed as $H=\sum_x H_x$ where
each term $H_x$ is either a connection $\sum_{i,j} t^x_{ij} A^x_i B^x_j$ or an on-site term
$\sum_{i} C^x_i $.
(This decomposition is
unique up to a canonical transformation, which would change accordingly the collapse basis, and thus, the decomposition
can be considered unique for our purposes.)
Therefore, a condition necessary for ergodicity is that no Hamiltonian term be diagonal in the collapse basis.

Collapse bases can but do not have to be completely random: Bases that do not mix states with different particle number or
bases  with the spins quantized in a given direction are examples.
It has also been suggested\cite{re:stoudenmire10} that changing the collapse basis
 from iteration to iteration (instead of keeping it fixed as has been assumed so far) improves ergodicity.

\subsection{Computational Complexity}

The computational complexity of the Krylov-space approach for the time evolution was mentioned in reference \onlinecite{re:alvarez11}:
The error dependence of the Krylov-space approach is\cite{re:manmana05} proportional to $\exp(-(\rho dt)^2/(16n)) (e\rho dt/(4n))^n$
with $n\ge \rho dt/2$ and $\rho$ the width of the spectrum, whereas in the Suzuki-Trotter approach the error is
given by the Trotter error.\cite{re:batrouni06}
The Krylov-space method is independent of the form of the Hamiltonian; the Suzuki-Trotter depends on the Hamiltonian connections.
The main disadvantage of the Krylov-space method is that it can be computationally more expensive compared to the Suzuki-Trotter
 due to the former requiring a tridiagonal decomposition
of the Hamiltonian. For the production of METTS, however, there is no ground state computation, and there is a single tridiagonal decomposition
needed per step. As noted in reference \onlinecite{re:white09}, the computational complexity
of the METTS algorithm is that of ground state DMRG multiplied by $\beta$ and by the number of measurements needed.

Parallelization was implemented in various places: in the Hamiltonian construction, in the construction of the density matrix, in
the wave function transformations, in the computation of two-point correlations.
Parallelization helps decrease the pre-factors in the CPU times, but does not affect the scaling in other ways.
The CPU time required for long chains was about 12 hours for 100 METTS. These CPU times
double if two-point correlations are needed. Also, sometimes multiple series of METTS need to be produced, as
will be explained in the next section.

\section{Results}\label{sec:results}
\subsection{Site-dependent potential}
Consider a one-orbital Hubbard model,
\begin{equation}
H=\sum_{i,j,\sigma}t_{ij} c^\dagger_{i\sigma}c_{j\sigma} + U\sum_i n_{i\uparrow}n_{i\downarrow} + \sum_{i,\sigma} V_i n_{i,\sigma},
\label{eq:ham}
\end{equation}
where $t_{ij}$ corresponds to an open chain.
First let us set $U=0$ to be able to
compare with the exact result.
For $N=8$ and a fixed potential profile,
the energy and density were obtained by averaging over METTS. Numerical values and statistics are shown in table \ref{tbl:statistics}
as a function of the inverse temperature $\beta$. Higher temperatures yield larger standard deviations, but the differences with the
exact results never exceed 8\% for the density, and 3\% for the energy. Longer runs can be performed to decrease these differences
even further if necessary.

\begin{figure}
\includegraphics[width=\mywidth,clip]{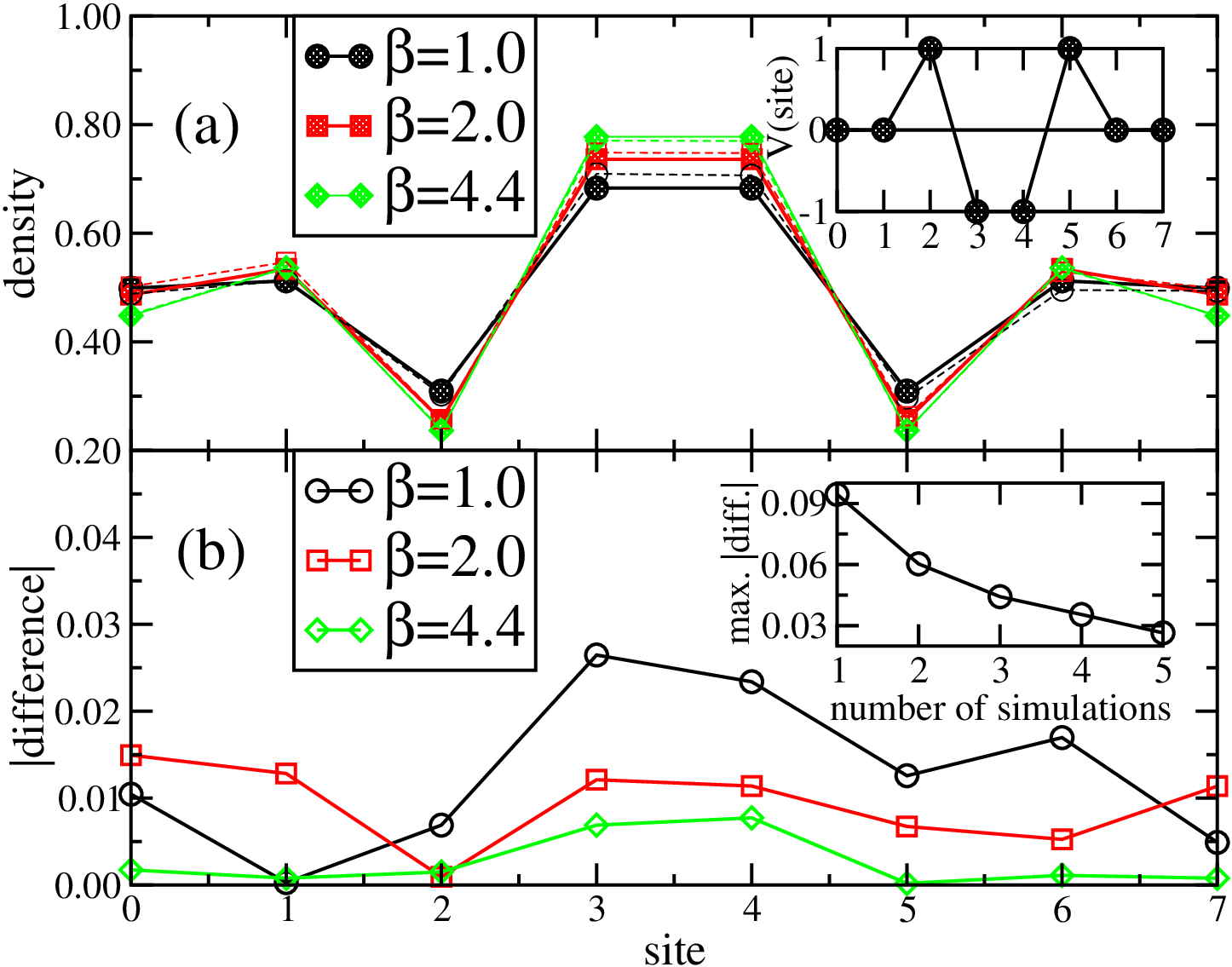}
\caption{(Color online) (a) Density at each site of the same system as in table \ref{tbl:statistics}, with $\beta$ as indicated.
Exact results are given by solid lines
and filled symbols, METTS results by dashed lines and open symbols. Inset: potential profile for this system.
(b) Unsigned difference between the density computed by the METTS algorithm and the exact one, at each site of the chain. Temperatures as before.
An average over 5 series of METTS was performed.
Inset: Maximum difference at $\beta=1.0$ as a function of the number of simulations, that is, the number of series of
METTS used.\label{fig:metts5Density}}
\end{figure}

Figure \ref{fig:metts5Density}(a) shows the resulting density profile, where only one spin sector is considered.
The observable $n_i$ differs from the exact result more than global observables like the total energy.
As mentioned, to decrease this difference longer runs could be performed, but it turns out to be better to produce
multiple series of METTS started from
different CPSs, and average them.
The latter strategy has the advantage of completely smearing the dependence on the initial configuration,
decreasing autocorrelation times faster.
Figure \ref{fig:metts5Density}(b) shows the density profile after averaging five series of METTS.
On the right, the inset of figure \ref{fig:metts5Density} shows the maximum difference between the METTS result and
the exact result at the highest temperature considered ($\beta=1.0$) as a function of the
number of series of METTS used, confirming the effectiveness of this method even for a small number of series.
\begin{table}
\begin{tabular}{|l|l|l|l|l|l|l|}\hline
$\beta$ & $\langle n\rangle$ & Std. Dev. & Exact & $\langle E\rangle$ & Std. Dev. & Exact\\\hline
1.0 & 3.900 & 0.550 & 4.007  & -7.1360 & 0.9715 & -6.9774\\\hline
2.0 & 4.278 & 0.362 &  4.027   & -9.0594 & 0.5785 & -9.3178  \\\hline
4.4 & 3.962 & 0.123 & 4.002  & -10.4584 & 0.2169 & -10.3515\\\hline
\end{tabular}
\caption{Mean values, standard deviations and exact results for both energy and density as a
function of the inverse temperature $\beta$ for an eight-site open chain with
$U=0$ and the potential profile shown in the inset of figure \ref{fig:metts5Density}(a).
Approximately 150 METTS were used.
\label{tbl:statistics}}
\end{table}

\subsection{Attractive Hubbard Model}

\subsubsection{Pairing correlation}

Let us now study Eq.~(\ref{eq:ham}) at quarter filling with $U<0$ and $V_i=\mu$ $\forall i$.
The density ($\langle n\rangle=0.25$) is \emph{fixed on average} only, and simulations are carried out in the grand
canonical ensemble in order to collapse to completely random bases. A particle- (but not $S_z$-) conserving basis
would be better here were it not for the ergodicity problems---which I have found to be too severe.
The choice of model, the attractive Hubbard model, is based on the following considerations.
The METTS algorithm has not been tested on fermionic systems before.
The Hubbard model on a periodic chain has been solved exactly,\cite{re:lieb68} albeit correlations are difficult to compute.
The attractive Hubbard model
superconducting correlations are known, and can be computed and extrapolated at moderate lattice sizes.

The s-wave pairing function
(see, for example, reference \onlinecite{re:guerrero00})
$P_s(R)=\frac 1N \sum_i \langle\Delta^\dagger_s(i+R)\Delta_s(i)\rangle$
measures superconducting correlations in this model; here $\Delta_s(i)=c_{i\downarrow}c_{i\uparrow}$.
We are going to consider $P_s(R)$
 as a function of temperature $T$ at quarter filling.

For $T=0$ and a \emph{periodic} system $P_s(R)$
 is known\cite{re:kawakami91}  to decay as a power law.

Studying this decay requires taking into account pairs of sites at distances as large as possible,
while at the same time minimizing effects due to border sites.
One possibility is to avoid some sites next to the left and right borders of the chain, and average over all other pairs
at a given distance. Another possibility is to take into account
all pairs of sites at distance $R$ but restrict $R$ to half the length of the chain, an approach that includes
as many lattice sites as possible and
avoids both sites in the pair being close to the borders for large distances. I have estimated and compared exponents of power laws
following both approaches, and found small differences but no change in the trends of the exponents.
Therefore, consistent with the second approach mentioned, distances $R>N/2$ were discarded in order to avoid boundary effects.

\begin{figure}
\includegraphics[width=\mywidth,clip]{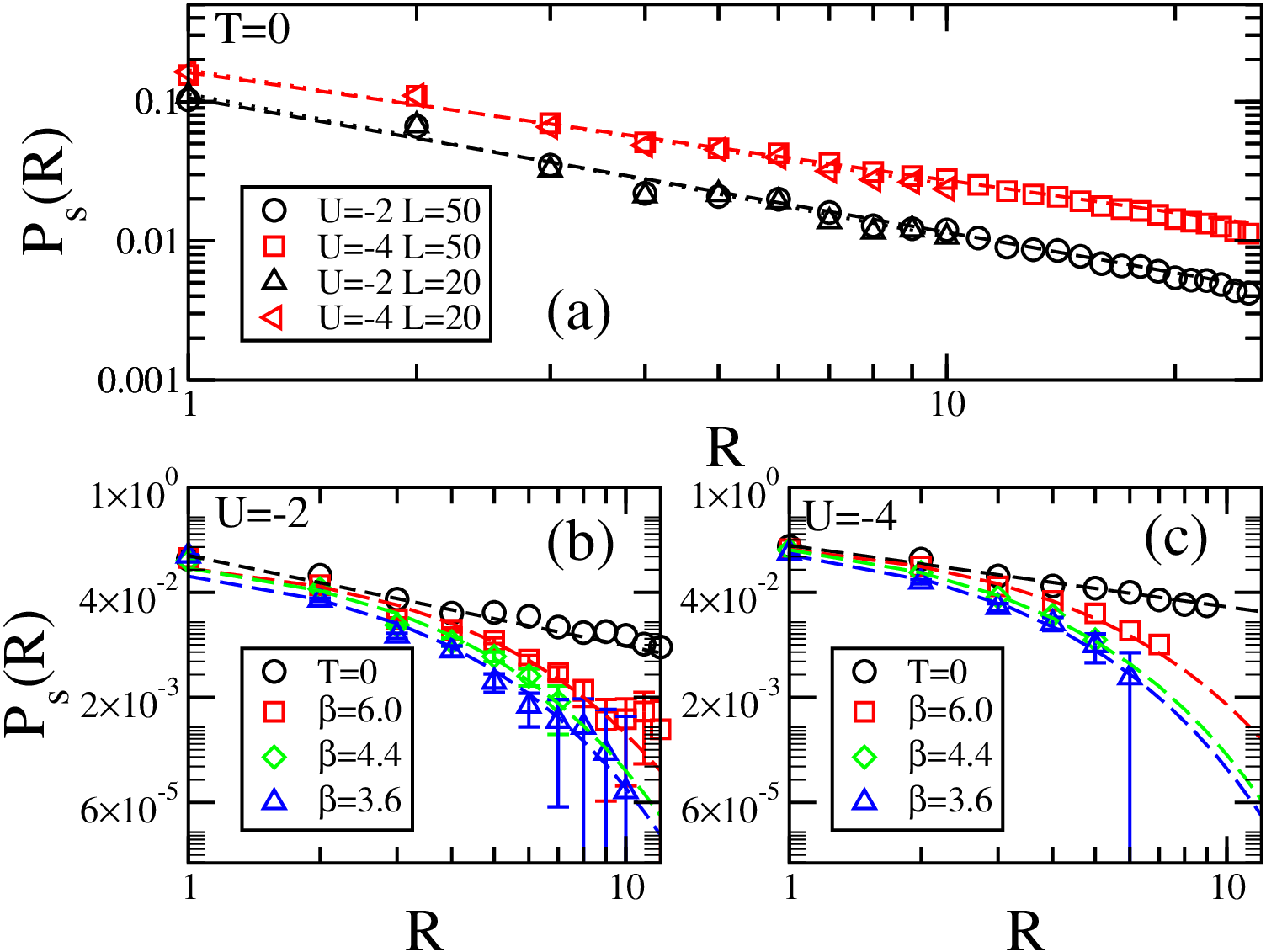}
\caption{(Color online) (a) $P_s(R)$ versus $R$ for different $U$ values and for two lattice sizes $N=50$ and $N=20$, always at quarter filling,
obtained from the ground state DMRG calculation.
(b) $P_s(R)$ versus $R$ for $U=-2$ for a chain with $N=20$ sites as a function of $\beta$. The $T=0$ values
scale as a power law. The $T>0$ values scale exponentially.
 Data points were deleted beyond a certain $R$; the text explains.
(c) Same as (b) for $U=-4$.
For the values of all exponents see table \ref{tbl:exponents}.
\label{fig:correlationT0}}
\end{figure}

Results are shown in figure \ref{fig:correlationT0}(a),
and exponents $\bar{\beta}$ for the expression $P_s(R)\propto R^{-\bar{\beta}}$ in columns 2 to 4
of table \ref{tbl:exponents}, exponents that turn out to be
comparable to those of Figure 8 of reference \onlinecite{re:kawakami91} at quarter filling.
(Although the standard notation for this exponent is $\beta$, the exponent is here denoted by $\bar{\beta}$
to avoid confusion with the inverse temperature.)
Trying to fit these $T=0$ results to an exponential always gives \emph{larger} errors than trying to fit them
to a power law.
For $T=0$ the power law exponents computed here with DMRG decrease with increasing $U$, as is the case
in the exact thermodynamic limit results.
A value of $m=200$ was used in this ground state DMRG calculation, where
each site of the lattice was swept three times.

For $T>0$ $P_s(R)$ is known\cite{re:koma92,re:tasaki98} to decay with exponential bounds.

Using the METTS algorithm to simulate finite temperature, a faster decay can be seen than that at $T=0$:
Results for $T>0$ in figure \ref{fig:correlationT0}(b) for $U=-2$ and figure \ref{fig:correlationT0}(c) for $U=-4$ depart
from the power law scaling (represented by a straight line in logarithmic scale), and approximate an exponential
scaling $P_s(R)\propto \exp(-\gamma R)$.
Exponents $\gamma$ for different values of $U$ are given in  columns 5 to 7
of table \ref{tbl:exponents}. For large $\beta$, $\gamma$ must be proportional to $1/\beta$,\cite{re:koma92,re:tasaki98} and
for $\beta\ge4.4$ this relation indeed holds with 3\% accuracy to 12\% accuracy depending on $U$, as shown in the last
column of table \ref{tbl:exponents}.
Trying to fit METTS results (at $T>0$) to an exponential always gives \emph{smaller} errors than trying to fit them
to a power law.
The exponential law exponents $\gamma$ decrease with $\beta$, as expected.\cite{re:koma92,re:tasaki98}.
At fixed $\beta$, the exponents first decrease from $U=-2$ to $U=-4$, and
then they increase from $U=-4$ to $U=-8$.

\begin{table}
\begin{tabular}{|l|l|l|l||l|l|l||l|}\hline
$U$ & $20\times1$ & $50\times1$& $d=1$& $\beta=6.0$&  $\beta=4.4$ & $\beta=3.6$ & \\\hline
-2 & 1.055 & 0.991 &  0.80  & 0.57 & 0.69 & 0.72 & 12\%\\\hline
-4 & 0.867 & 0.784 & 0.70 & 0.48  & 0.68 & 0.72 &3\%\\\hline
-8 & 0.808 & 0.725 & 0.65 & 0.94 & 1.21  & 1.34 & 6\%\\\hline
\end{tabular}
\caption{Columns 2 and 3 show exponents $\bar{\beta}$ for the power law fit done  at $T=0$ in
 figure \ref{fig:correlationT0}(a).
Values of $\bar{\beta}$ at the thermodynamic limit in one dimension are indicated by $d=1$, and obtained from  Figure 8 of
reference \onlinecite{re:kawakami91} at quarter filling.
The next three columns show exponents $\gamma$ for the exponential fit done at $T>0$ for $N=20$ in
figures \ref{fig:correlationT0}(b) and (c).
The last column is $100\times|1-6.0\times\gamma(6.0)/(4.4\times\gamma(4.4)|$.
\label{tbl:exponents}}
\end{table}

Several caveats regarding the way exponents were obtained need to be noted.
In these METTS simulations $\mu$ had to be fixed to yield a density  approximate to
quarter filling. Actual densities obtained were in the range 0.2 to 0.3
instead of exactly at 0.25.
$P_s(R)$ at finite and, in particular, large temperatures
has large statistical errors. The first five METTS were discarded due to convergence reasons, similar to the case in figure~\ref{fig:figure1mu}.
Approximately 10 to 20 METTS were used for measurements, and multiple METTS series run starting from different and random CPSs.
Most results were computed with a fixed and random collapse basis. A few controls were performed with computational runs with
the bases changed at each collapse; discrepancies between the two collapse procedures fell within the error bars.
For the purposes of fitting to an exponential,
values beyond a certain $R$ were discarded when either (i) behavior became non-monotonic, or (ii) $P_s(R)$ became slightly negative.
These two effects are due to statistical errors: In most cases, discarded points
had errors---shown by the error bars in figure \ref{fig:correlationT0}(b,c)---larger than their average values.

\subsubsection{Non-local correlation}

\begin{figure}
\includegraphics[width=\mywidth,clip]{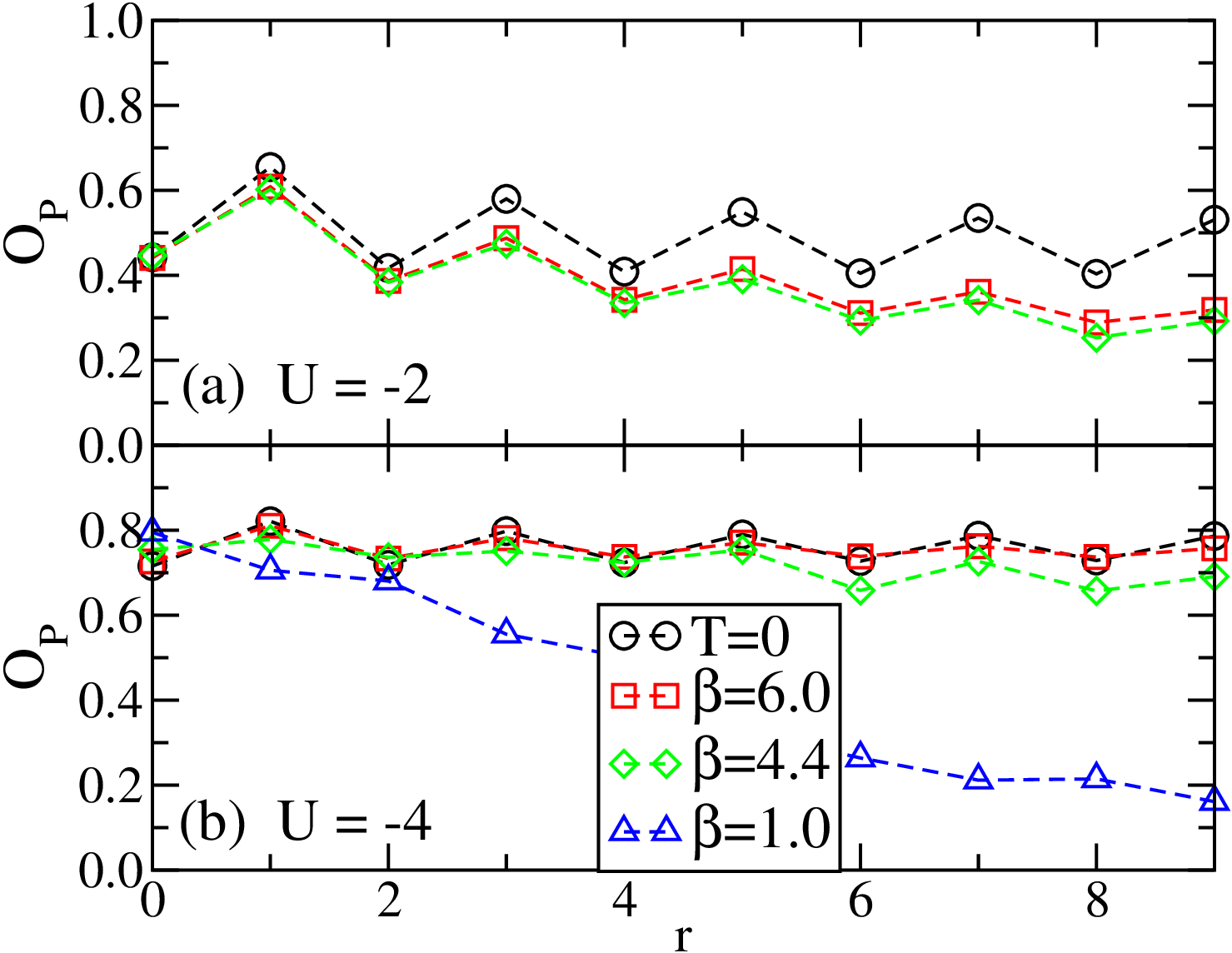}
\caption{(Color online) $O_P^{(s)}(r)$ vs $r$ for chains of 20 sites at half filling and (a) $U=-2$ and (b) $U=-4$.
\label{fig:parityCorrelator}}
\end{figure}

Let us now consider the non-local parity correlator for charge
\begin{equation}
O_P^{(s)}(r) = \langle \exp(2i\pi\sum_{j=i}^{i+r} S^z_{j})\rangle,
\end{equation}
with $S^z_i=\frac 12 (n_{i\uparrow}-n_{i\downarrow})$, which is discussed in \onlinecite{re:montorsi12} and references therein.
Figure \ref{fig:parityCorrelator} shows $O_P^{(s)}(r)$ vs $r$ at $T=0$ and at $T>0$, for chains of 20 sites at half filling,
and for two values of $U$, as indicated.

In the $U<0$ case considered here the Luther-Emery phase is characterized by nonzero $O_P^{(s)}(r)$,\cite{re:montorsi12} and
the transition is known
to be of Berezinskii-Kosterlitz-Thouless (BKT) type.
The non-local order parameter $O_P^{(s)}(r)$ remains finite even at $T>0$, as seen in the figure.

\section{Summary and Outlook}\label{sec:summary}

A procedural description for the generation of METTS was presented, using the Krylov-space approach to perform the
imaginary time evolution of the classical product states.
The full open source code, input decks and additional computational details have been made available
at \url{https://web.ornl.gov/~gz1/papers/40/}.\cite{re:ince12}
By averaging over 100 to 200 METTS, global observables can be obtained with reasonable errors.
Applied to local observables, in the example of the density in a site-dependent potential,
the simulations converge to exact results as more series of METTS are added.

For an attractive Hubbard model on a chain,
the application of the METTS algorithm  verified the exponential decay of correlations---as opposed to the power law
decay for the ground state. Or, conversely, the correct behavior of the exponents obtained with the METTS algorithm
verified its feasibility. The exponents $\gamma$ are proportional to the temperature for low enough temperatures,
as expected from rigorous bounds. A similar departure from power law at finite temperature
has been observed for a spinless model with nearest and next-nearest neighbor interactions.\cite{re:karrasch12}

These studies should set the stage for the further use of METTS in fermionic systems. We should envision  using
METTS to compute transition temperatures in models of pnictides superconductors, where, as mentioned before,
spin flipping terms might preclude the use of quantum Monte Carlo methods due to the sign problem.
One difficulty will be the vanishing of superconducting correlations in two dimensional
systems,\cite{re:mermin66,re:kosterlitz73,re:paiva04} a difficulty that
could be overcome by
explicitly breaking symmetries with an anisotropic term.\cite{re:gelfert01}
Multiple orbital models, as needed for pnictides superconductors, have larger Hilbert spaces, and would require longer runs.
Let us not forget, however, that estimating exponents (as was done in this work) has high demands in terms of accuracy.
But for multi-orbital superconductors, computing transition temperatures would be the main interest and motivation, and
computing transition temperatures---assuming a true second order transition is present---could be achieved with less measurements than needed for
estimating exponents.

\begin{acknowledgments}
I would like to thank K.~Al-Hassanieh, T.~Maier, J.~Rinc\'on, E.~M.~Stoudenmire and S.~R.~White for helpful discussions and suggestions.
This research was conducted at the
Center for Nanophase Materials Sciences at Oak Ridge National Laboratory,
sponsored by the Scientific User Facilities Division, Basic Energy Sciences, U.S.
Department of Energy (DOE),
under contract with UT-Battelle.
I would like to acknowledge support from the DOE early career research program.
\end{acknowledgments}

\bibliography{thesis}

\begin{thebibliography}{30}
\expandafter\ifx\csname natexlab\endcsname\relax\def\natexlab#1{#1}\fi
\expandafter\ifx\csname bibnamefont\endcsname\relax
  \def\bibnamefont#1{#1}\fi
\expandafter\ifx\csname bibfnamefont\endcsname\relax
  \def\bibfnamefont#1{#1}\fi
\expandafter\ifx\csname citenamefont\endcsname\relax
  \def\citenamefont#1{#1}\fi
\expandafter\ifx\csname url\endcsname\relax
  \def\url#1{\texttt{#1}}\fi
\expandafter\ifx\csname urlprefix\endcsname\relax\def\urlprefix{URL }\fi
\providecommand{\bibinfo}[2]{#2}
\providecommand{\eprint}[2][]{\url{#2}}

\bibitem[{\citenamefont{White}(1992)}]{re:white92}
\bibinfo{author}{\bibfnamefont{S.}~\bibnamefont{White}},
  \bibinfo{journal}{Phys. Rev. Lett.} \textbf{\bibinfo{volume}{69}},
  \bibinfo{pages}{2863} (\bibinfo{year}{1992}).

\bibitem[{\citenamefont{White}(1993)}]{re:white93}
\bibinfo{author}{\bibfnamefont{S.}~\bibnamefont{White}},
  \bibinfo{journal}{Phys. Rev. B} \textbf{\bibinfo{volume}{48}},
  \bibinfo{pages}{345} (\bibinfo{year}{1993}).

\bibitem[{\citenamefont{Verstraete et~al.}(2004)\citenamefont{Verstraete,
  Garcia-Ripoll, and Cirac}}]{re:verstraete04}
\bibinfo{author}{\bibfnamefont{F.}~\bibnamefont{Verstraete}},
  \bibinfo{author}{\bibfnamefont{J.~J.} \bibnamefont{Garcia-Ripoll}},
  \bibnamefont{and} \bibinfo{author}{\bibfnamefont{J.~I.} \bibnamefont{Cirac}},
  \bibinfo{journal}{Phys. Rev. Lett.} \textbf{\bibinfo{volume}{93}},
  \bibinfo{pages}{207204} (\bibinfo{year}{2004}).

\bibitem[{\citenamefont{Zwolak and Vidal}(2004)}]{re:zwolak04}
\bibinfo{author}{\bibfnamefont{M.}~\bibnamefont{Zwolak}} \bibnamefont{and}
  \bibinfo{author}{\bibfnamefont{G.}~\bibnamefont{Vidal}},
  \bibinfo{journal}{Phys. Rev. Lett.} \textbf{\bibinfo{volume}{93}},
  \bibinfo{pages}{207205} (\bibinfo{year}{2004}).

\bibitem[{\citenamefont{Feiguin and White}(2005)}]{re:feiguin05}
\bibinfo{author}{\bibfnamefont{A.~R.} \bibnamefont{Feiguin}} \bibnamefont{and}
  \bibinfo{author}{\bibfnamefont{S.~R.} \bibnamefont{White}},
  \bibinfo{journal}{Phys. Rev. B} \textbf{\bibinfo{volume}{72}},
  \bibinfo{pages}{020404} (\bibinfo{year}{2005}).

\bibitem[{\citenamefont{White}(2009)}]{re:white09}
\bibinfo{author}{\bibfnamefont{S.}~\bibnamefont{White}},
  \bibinfo{journal}{Phys. Rev. Lett.} \textbf{\bibinfo{volume}{102}},
  \bibinfo{pages}{190601} (\bibinfo{year}{2009}).

\bibitem[{\citenamefont{Stoudenmire and White}(2010)}]{re:stoudenmire10}
\bibinfo{author}{\bibfnamefont{E.}~\bibnamefont{Stoudenmire}} \bibnamefont{and}
  \bibinfo{author}{\bibfnamefont{S.}~\bibnamefont{White}},
  \bibinfo{journal}{New J. Phys.} \textbf{\bibinfo{volume}{12}},
  \bibinfo{pages}{055026} (\bibinfo{year}{2010}).

\bibitem[{\citenamefont{Scholl{w\"o}ck}(2010)}]{re:schollwock10}
\bibinfo{author}{\bibfnamefont{U.}~\bibnamefont{Scholl{w\"o}ck}},
  \bibinfo{journal}{Annals of Physics} \textbf{\bibinfo{volume}{96}},
  \bibinfo{pages}{326} (\bibinfo{year}{2010}).

\bibitem[{\citenamefont{Schollw{\"o}ck}(2009)}]{re:schollwock09}
\bibinfo{author}{\bibfnamefont{U.}~\bibnamefont{Schollw{\"o}ck}},
  \bibinfo{journal}{Physics} \textbf{\bibinfo{volume}{2}}, \bibinfo{pages}{39}
  (\bibinfo{year}{2009}).

\bibitem[{\citenamefont{Kamihara et~al.}(2008)\citenamefont{Kamihara, Watanabe,
  Hirano, and Hosono}}]{re:kamihara08}
\bibinfo{author}{\bibfnamefont{Y.}~\bibnamefont{Kamihara}},
  \bibinfo{author}{\bibfnamefont{T.}~\bibnamefont{Watanabe}},
  \bibinfo{author}{\bibfnamefont{M.}~\bibnamefont{Hirano}}, \bibnamefont{and}
  \bibinfo{author}{\bibfnamefont{H.}~\bibnamefont{Hosono}},
  \bibinfo{journal}{J. Am Chem. Soc.} \textbf{\bibinfo{volume}{130}},
  \bibinfo{pages}{3296} (\bibinfo{year}{2008}).

\bibitem[{\citenamefont{Guo et~al.}(2005)\citenamefont{Guo, Jin, Wang, Wang,
  Zhu, Zhou, He, and Chen}}]{re:guo10}
\bibinfo{author}{\bibfnamefont{J.}~\bibnamefont{Guo}},
  \bibinfo{author}{\bibfnamefont{S.}~\bibnamefont{Jin}},
  \bibinfo{author}{\bibfnamefont{G.}~\bibnamefont{Wang}},
  \bibinfo{author}{\bibfnamefont{S.}~\bibnamefont{Wang}},
  \bibinfo{author}{\bibfnamefont{K.}~\bibnamefont{Zhu}},
  \bibinfo{author}{\bibfnamefont{T.}~\bibnamefont{Zhou}},
  \bibinfo{author}{\bibfnamefont{M.}~\bibnamefont{He}}, \bibnamefont{and}
  \bibinfo{author}{\bibfnamefont{X.}~\bibnamefont{Chen}},
  \bibinfo{journal}{Phys. Rev. B} \textbf{\bibinfo{volume}{82}},
  \bibinfo{pages}{180520(R)} (\bibinfo{year}{2005}).

\bibitem[{\citenamefont{Daghofer et~al.}(2008)\citenamefont{Daghofer, Moreo,
  Riera, Arrigoni, Scalapino, and Dagotto}}]{re:daghofer08}
\bibinfo{author}{\bibfnamefont{M.}~\bibnamefont{Daghofer}},
  \bibinfo{author}{\bibfnamefont{A.}~\bibnamefont{Moreo}},
  \bibinfo{author}{\bibfnamefont{J.~A.} \bibnamefont{Riera}},
  \bibinfo{author}{\bibfnamefont{E.}~\bibnamefont{Arrigoni}},
  \bibinfo{author}{\bibfnamefont{D.}~\bibnamefont{Scalapino}},
  \bibnamefont{and} \bibinfo{author}{\bibfnamefont{E.}~\bibnamefont{Dagotto}},
  \bibinfo{journal}{Phys. Rev. Lett.} \textbf{\bibinfo{volume}{101}},
  \bibinfo{pages}{237004} (\bibinfo{year}{2008}).

\bibitem[{\citenamefont{Moreo et~al.}(2008)\citenamefont{Moreo, Daghofer, and
  Dagotto}}]{re:moreo08}
\bibinfo{author}{\bibfnamefont{A.}~\bibnamefont{Moreo}},
  \bibinfo{author}{\bibfnamefont{M.}~\bibnamefont{Daghofer}}, \bibnamefont{and}
  \bibinfo{author}{\bibfnamefont{E.}~\bibnamefont{Dagotto}},
  \bibinfo{journal}{Phys. Rev. B} \textbf{\bibinfo{volume}{79}},
  \bibinfo{pages}{104510} (\bibinfo{year}{2008}).

\bibitem[{\citenamefont{Manmana et~al.}(2005)\citenamefont{Manmana, Muramatsu,
  and Noack}}]{re:manmana05}
\bibinfo{author}{\bibfnamefont{S.~R.} \bibnamefont{Manmana}},
  \bibinfo{author}{\bibfnamefont{A.}~\bibnamefont{Muramatsu}},
  \bibnamefont{and} \bibinfo{author}{\bibfnamefont{R.~M.} \bibnamefont{Noack}},
  in \emph{\bibinfo{booktitle}{AIP Conf. Proc.}} (\bibinfo{year}{2005}), vol.
  \bibinfo{volume}{789}, pp. \bibinfo{pages}{269--278}, \bibinfo{note}{also in
  http://arxiv.org/abs/cond-mat/0502396v1}.

\bibitem[{\citenamefont{Schollw{\"o}eck and White}(2006)}]{re:batrouni06}
\bibinfo{author}{\bibnamefont{Schollw{\"o}eck}} \bibnamefont{and}
  \bibinfo{author}{\bibnamefont{White}}, in \emph{\bibinfo{booktitle}{Effective
  models for low-dimensional strongly correlated systems}}, edited by
  \bibinfo{editor}{\bibfnamefont{G.~G.} \bibnamefont{Batrouni}}
  \bibnamefont{and} \bibinfo{editor}{\bibfnamefont{D.}~\bibnamefont{Poilblanc}}
  (\bibinfo{publisher}{AIP}, \bibinfo{address}{Melville, New York},
  \bibinfo{year}{2006}), p. \bibinfo{pages}{155}, \bibinfo{note}{also in
  http://de.arxiv.org/abs/cond-mat/0606018v1}.

\bibitem[{\citenamefont{Schollw{\"o}ck}(2005)}]{re:schollwock05}
\bibinfo{author}{\bibfnamefont{U.}~\bibnamefont{Schollw{\"o}ck}},
  \bibinfo{journal}{Rev. Mod. Phys.} \textbf{\bibinfo{volume}{77}},
  \bibinfo{pages}{259} (\bibinfo{year}{2005}).

\bibitem[{\citenamefont{Alvarez et~al.}(2011)\citenamefont{Alvarez, da~Silva,
  Ponce, and Dagotto}}]{re:alvarez11}
\bibinfo{author}{\bibfnamefont{G.}~\bibnamefont{Alvarez}},
  \bibinfo{author}{\bibfnamefont{L.~G. G. V.~D.} \bibnamefont{da~Silva}},
  \bibinfo{author}{\bibfnamefont{E.}~\bibnamefont{Ponce}}, \bibnamefont{and}
  \bibinfo{author}{\bibfnamefont{E.}~\bibnamefont{Dagotto}},
  \bibinfo{journal}{Phys. Rev. E} \textbf{\bibinfo{volume}{84}},
  \bibinfo{pages}{056706} (\bibinfo{year}{2011}).

\bibitem[{\citenamefont{Lieb and Wu}(1968)}]{re:lieb68}
\bibinfo{author}{\bibfnamefont{E.~H.} \bibnamefont{Lieb}} \bibnamefont{and}
  \bibinfo{author}{\bibfnamefont{F.~Y.} \bibnamefont{Wu}},
  \bibinfo{journal}{Phys. Rev. Lett.} \textbf{\bibinfo{volume}{20}},
  \bibinfo{pages}{1445} (\bibinfo{year}{1968}).

\bibitem[{\citenamefont{Guerrero et~al.}(2000)\citenamefont{Guerrero, Ortiz,
  and Gubernatis}}]{re:guerrero00}
\bibinfo{author}{\bibfnamefont{M.}~\bibnamefont{Guerrero}},
  \bibinfo{author}{\bibfnamefont{G.}~\bibnamefont{Ortiz}}, \bibnamefont{and}
  \bibinfo{author}{\bibfnamefont{J.~E.} \bibnamefont{Gubernatis}},
  \bibinfo{journal}{Phys. Rev. B} \textbf{\bibinfo{volume}{62}},
  \bibinfo{pages}{600} (\bibinfo{year}{2000}).

\bibitem[{\citenamefont{Kawakami and Yang}(1991)}]{re:kawakami91}
\bibinfo{author}{\bibfnamefont{N.}~\bibnamefont{Kawakami}} \bibnamefont{and}
  \bibinfo{author}{\bibfnamefont{S.-K.} \bibnamefont{Yang}},
  \bibinfo{journal}{Phys. Rev. B} \textbf{\bibinfo{volume}{44}},
  \bibinfo{pages}{7844} (\bibinfo{year}{1991}).

\bibitem[{\citenamefont{Koma and Tasaki}(1992)}]{re:koma92}
\bibinfo{author}{\bibfnamefont{T.}~\bibnamefont{Koma}} \bibnamefont{and}
  \bibinfo{author}{\bibfnamefont{H.}~\bibnamefont{Tasaki}},
  \bibinfo{journal}{Phys. Rev. Lett.} \textbf{\bibinfo{volume}{68}},
  \bibinfo{pages}{3248} (\bibinfo{year}{1992}).

\bibitem[{\citenamefont{Tasaki}(1998)}]{re:tasaki98}
\bibinfo{author}{\bibfnamefont{H.}~\bibnamefont{Tasaki}}, \bibinfo{journal}{J.
  Phys.: Condens. Matter} \textbf{\bibinfo{volume}{10}}, \bibinfo{pages}{4353}
  (\bibinfo{year}{1998}).

\bibitem[{\citenamefont{Montorsi and Roncaglia}(͑2012)}]{re:montorsi12}
\bibinfo{author}{\bibfnamefont{A.}~\bibnamefont{Montorsi}} \bibnamefont{and}
  \bibinfo{author}{\bibfnamefont{M.}~\bibnamefont{Roncaglia}},
  \bibinfo{journal}{Phys. Rev. Lett.} \textbf{\bibinfo{volume}{109}},
  \bibinfo{pages}{236404} (\bibinfo{year}{͑2012}).

\bibitem[{re:()}]{re:note2013a}
\bibinfo{note}{{S}ee Supplemental Material at [URL will be inserted by
  publisher] for a more detailed description of the numerical data shown.}

\bibitem[{\citenamefont{Ince et~al.}(2012)\citenamefont{Ince, Hatton, and
  Graham-Cumming}}]{re:ince12}
\bibinfo{author}{\bibfnamefont{D.~C.} \bibnamefont{Ince}},
  \bibinfo{author}{\bibfnamefont{L.}~\bibnamefont{Hatton}}, \bibnamefont{and}
  \bibinfo{author}{\bibfnamefont{J.}~\bibnamefont{Graham-Cumming}},
  \bibinfo{journal}{Nature} \textbf{\bibinfo{volume}{482}},
  \bibinfo{pages}{485} (\bibinfo{year}{2012}).

\bibitem[{\citenamefont{Karrasch and Moore}(2012)}]{re:karrasch12}
\bibinfo{author}{\bibfnamefont{C.}~\bibnamefont{Karrasch}} \bibnamefont{and}
  \bibinfo{author}{\bibfnamefont{J.~E.} \bibnamefont{Moore}},
  \bibinfo{journal}{Phys. Rev. B} \textbf{\bibinfo{volume}{86}},
  \bibinfo{pages}{155156} (\bibinfo{year}{2012}).

\bibitem[{\citenamefont{Mermin and Wagner}(1966)}]{re:mermin66}
\bibinfo{author}{\bibfnamefont{N.~D.} \bibnamefont{Mermin}} \bibnamefont{and}
  \bibinfo{author}{\bibfnamefont{H.}~\bibnamefont{Wagner}},
  \bibinfo{journal}{Phys. Rev. Lett.} \textbf{\bibinfo{volume}{17}},
  \bibinfo{pages}{1133} (\bibinfo{year}{1966}).

\bibitem[{\citenamefont{Kosterlitz and Thouless}(͑1973͒)}]{re:kosterlitz73}
\bibinfo{author}{\bibfnamefont{J.}~\bibnamefont{Kosterlitz}} \bibnamefont{and}
  \bibinfo{author}{\bibfnamefont{D.}~\bibnamefont{Thouless}},
  \bibinfo{journal}{J. Phys. C} \textbf{\bibinfo{volume}{6}},
  \bibinfo{pages}{1181} (\bibinfo{year}{͑1973͒}).

\bibitem[{\citenamefont{Paiva et~al.}(2004)\citenamefont{Paiva, dos Santos,
  Scalettar, and Denteneer}}]{re:paiva04}
\bibinfo{author}{\bibfnamefont{T.}~\bibnamefont{Paiva}},
  \bibinfo{author}{\bibfnamefont{R.~R.} \bibnamefont{dos Santos}},
  \bibinfo{author}{\bibfnamefont{R.~T.} \bibnamefont{Scalettar}},
  \bibnamefont{and} \bibinfo{author}{\bibfnamefont{P.~J.~H.}
  \bibnamefont{Denteneer}}, \bibinfo{journal}{Phys. Rev. B}
  \textbf{\bibinfo{volume}{69}}, \bibinfo{pages}{184501}
  (\bibinfo{year}{2004}).

\bibitem[{\citenamefont{Gelfert and Nolting}(2001)}]{re:gelfert01}
\bibinfo{author}{\bibfnamefont{A.}~\bibnamefont{Gelfert}} \bibnamefont{and}
  \bibinfo{author}{\bibfnamefont{W.}~\bibnamefont{Nolting}},
  \bibinfo{journal}{J. Phys.: Condens. Matter} \textbf{\bibinfo{volume}{13}},
  \bibinfo{pages}{R505} (\bibinfo{year}{2001}).

\end{thebibliography}

\end{document}